\documentclass[reprint,aps,prl,superscriptaddress,twocolumn,floatfix]{revtex4-1}
\usepackage{graphicx}
\usepackage{amsmath}
\usepackage{bm}
\usepackage{color}
\usepackage{soul}
\usepackage{ulem}

\begin{document}

\title{Density waves at the interface of a binary complex plasma}
\author{Li Yang}
\affiliation{College of Science, Donghua University, 201620 Shanghai, PR China}
\author{Mierk Schwabe}
\author{Sergey Zhdanov}
\author{Hubertus M. Thomas}
\affiliation{Institut f\"ur Materialphysik im Weltraum, Deutsches Zentrum f\"ur Luft- und Raumfahrt (DLR), 82234 We{\ss}ling, Germany}
\author{Andrey M Lipaev}
\author{Vladimir I Molotkov}
\author{Vladimir E Fortov}
\affiliation{Joint Institute for High Temperatures, 125412 Moscow, Russia}
\author{Jing Zhang}
\author{Cheng-Ran Du}
\email{chengran.du@dhu.edu.cn}
\affiliation{College of Science, Donghua University, 201620 Shanghai, PR China}

\begin{abstract}
Density waves were studied in a phase-separated binary complex plasma under microgravity conditions. For the big particles, waves were self-excited by the two-stream instability, while for small particles, they were excited by heartbeat instability with the presence of reversed propagating pulses of a different frequency. By studying the dynamics of wave crests at the interface, we recognize a ``collision zone'' and a ``merger zone'' before and after the interface, respectively. The results provide a generic picture of wave-wave interaction at the interface between two ``mediums''.
\end{abstract}

\maketitle

Wave behavior at the interface consists of various interesting phenomena including transmission, reflection and refraction, etc. \cite{Lekner} The early study of refractions can be dated back to the year $984$, when Snell's law was first proposed by a scientist, Ibn Sahl, in Baghdad \cite{Rashed:1990}. Up-to-date studies of waves at an interface include not only light waves \cite{Xiao:2010}, but also acoustic waves in solids \cite{Wen:2009} and fluids \cite{Godin:2006}, capillary waves in liquids \cite{Blanchette:2006}, electromagnetic waves \cite{Bertin:2012}, charge density waves \cite{Weitering:1999}, and spin waves \cite{Kajiwara:2010}. These studies result in wide applications such as dental diagnostics \cite{Lees:1968} and study of seismic waves in geoscience \cite{Calvert:2004}. Recent progresses in granular matter \cite{Nesterenko:2005} and colloidal physics \cite{Hernandez:2009} enable us to study the wave behavior at an interface with resolution of individual particles. However, rigid contacts in dense granular matters and strong dissipation in solvent in colloids prohibit studies at the kinetic level.

Dust density waves in complex plasmas provide a unique opportunity to study various aspects of wave propagation. A complex plasma is a weekly ionized gas containing electrons, ions, neutral atoms and small macroscopic particles \cite{Fortov:2005,Morfill:2009}. Such a system allows experimental studies of various physical processes occurring in liquids and solids at the kinetic level \cite{Chaudhuri:2011}. Since the first observation of self-excited dust acoustic waves, dust density waves have drawn much attention \cite{Rao:1990,Barkan:1995,Tanna:1996,Merlino:2014}. Streaming ions generate a Buneman-type instability \cite{Rosenberg:1996}. Thus, dust density waves can be self-excited if internal sources of free energy exist \cite{Barkan:1995}. They can also be triggered by the heartbeat instability \cite{Heidemann:2011,Mikikian:2007}. The particle dynamics can be recorded by video microscopy \cite{Schwabe:2007}. While some properties of waves such as growth and clustering can be studied by recording the dynamics of wave crests alone \cite{Thomas:2006,Menzel:2010,Williams:2014,Heinrich:2012}, tracking individual particles in waves reveals the wave-particle dynamics in much detail \cite{Schwabe:2007,Teng:2009,Tsai:2016}. Apart from self-excited waves, external excitation can also sustain continuous waves \cite{Schwabe:2008,Williams:2008} or trigger solitary waves in complex plasmas \cite{Bandyopadhyay:2008,Jaiswal:2016}.

A complex plasma consisting of two differently sized microparticles is known as binary complex plasma. Under certain conditions, two types of particles can be mixed and form a glassy system \cite{Du:2016}. Other phenomena such as phase separation \cite{Ivlev:2009,Wysocki:2010} and lane formation \cite{Suetterlin:2009} can also be studied in such systems. Recently it was discovered that phase separation can still occur due to the imbalance of forces under microgravity conditions despite the criteria of spinodal decomposition not being fulfilled \cite{Killer:2016}. An interface between two different types of particles emerges in such binary systems.

\begin{figure}[!ht]
\centerline{\includegraphics[bb=0 20 500 480, width=0.45\textwidth]{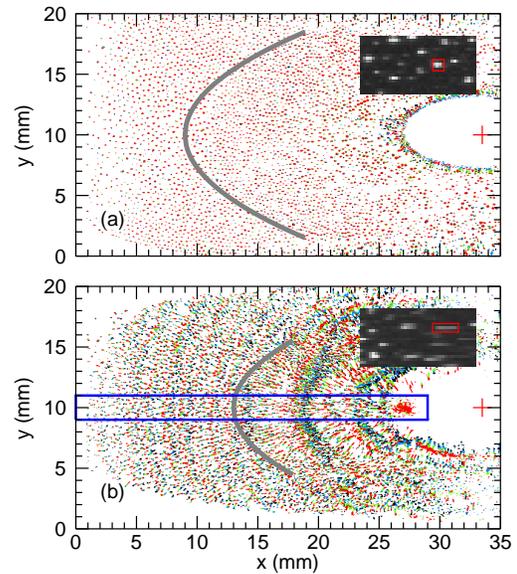}}
\caption{(Color online) Snapshot of a binary complex plasma at gas pressure $20$~Pa (a) and $10$~Pa (b). Five consecutive images are overlaid with different colors from blue to red. The interface is highlighted by a grey curve, where big particles are confined in the left area and small particles in the right area. The ROI for the periodgram in Fig.~\ref{figure2} is marked in blue. Because of high velocity, microparticle images have an elongated shape, shown in the raw image in the insets. The center of the vacuum chamber is marked by a red cross.
\label{figure1}}
\end{figure}

Despite plenty of studies on dust acoustic waves in past years, it is not clear how two waves of different origins interact at an interface. In this letter, we study the density waves at the interface of a phase-separated binary complex plasma under microgravity conditions. The emergence of two frequencies due to different excitation origins and respective features in terms of kinetics of individual particles are reported. A detailed study of the periodgram reveals characteristic zones in the vicinity of the interface where waves in two ``mediums'' strongly interact.

The experiments were performed in the PK-3~Plus laboratory on board the International Space Station (ISS). Technical details of the setup can be found in \cite{Thomas:2008,Schwabe:2008}. An argon plasma was produced by a capacitively-coupled radio-frequency (rf) generator in push-pull mode at $13.56$~Hz. We prepared a binary complex plasma by sequentially injecting melamine formaldehyde (MF) microparticles of two different sizes in the plasma discharge. The mass density of the microparticles is $1.51$~g/cm$^3$. The small particles have a diameter $d_s=6.8$~$\mu$m while the big ones have a diameter $d_b=9.2$~$\mu$m. With video microscopy \cite{Thomas:2008}, a cross section of the left half of the particle cloud (illuminated by a laser sheet) was recorded with a frame rate of $50$~frames-per-second (fps) and a spatial resolution of $0.05$~mm/pixel.

As we see in Fig.~\ref{figure1}(a), initially at a pressure of $20$~Pa, microparticles formed a 3-dimensional (3D) cloud with a particle-free region in the center (void) \cite{Morfill:1999,Avinash:2003}. Two particle species were phase-separated with a clear interface (highlighted by a grey curve) due to the following two mechanisms: First, the disparity of particle size (${\Delta}d/\bar{d} \approx 0.3$) was larger than the critical value of spinodal decomposition ($0.25$) \cite{Ivlev:2009}. Second, both particle species were subjected to two forces under microgravity conditions, namely ion drag force (directed outwards \footnote{The ions flow outwards from the cloud center (marked by a cross in Fig.~\ref{figure1}) on average, namely from right to left in Fig.~\ref{figure1} \cite{Thomas:2008}.}) and the electric field force (directed inwards). The total force acting on two particle species had a subtle difference depending on the particle diameter \cite{Killer:2016}. The synergistic effect of spinodal decomposition and force difference led to the instantaneous phase separation. Particularly the second effect drove the small particles into the inner part of the particle cloud and left big particles outside, as shown in Fig.~\ref{figure1}(a).

\begin{table*}[t]
\caption{Parameters including average number density $\langle n \rangle$, charge $Q$ and sound speed $C$ estimated for small and big particles. Among these parameters, the charge and the sound speed are estimated based on three theoretic models: The drift motion limited theory (DML) \cite{Morfill:2006}, the modified orbital motion limited theory (mOML) \cite{Khrapak:2005} and the orbital motion limited theory (OML) \cite{Bonitz:2010}.}
\label{table1}
\begin{center}
\begin{tabular}{cccccccc}
    \hline
    $d$ & $\langle n \rangle$ & $Q_{DML}$ & $Q_{mOML}$ & $Q_{OML}$ & $C_{DML}$  & $C_{mOML}$  & $C_{OML}$ \\
    $[\mu$m$]$ & $[$mm$^{-3}]$ & $[10^4e]$ & $[10^4e]$ & $[10^4e]$ & $[$mm/s$]$ & $[$mm/s$]$ & $[$mm/s$]$ \\
    \hline \hline
    $6.8$ & $240$ & $1.0$ & $0.9$ & $1.4$ & $10.5$ & $9.5$ & $12.9$ \\
    $9.2$ & $60$ & $1.7$ & $1.4$ & $1.9$ & $7.0$ & $5.9$ & $7.6$ \\
    \hline
\end{tabular}
\end{center}
\end{table*}

\begin{figure}[!ht]
\centerline{\includegraphics[bb=0 20 500 600, width=0.4\textwidth]{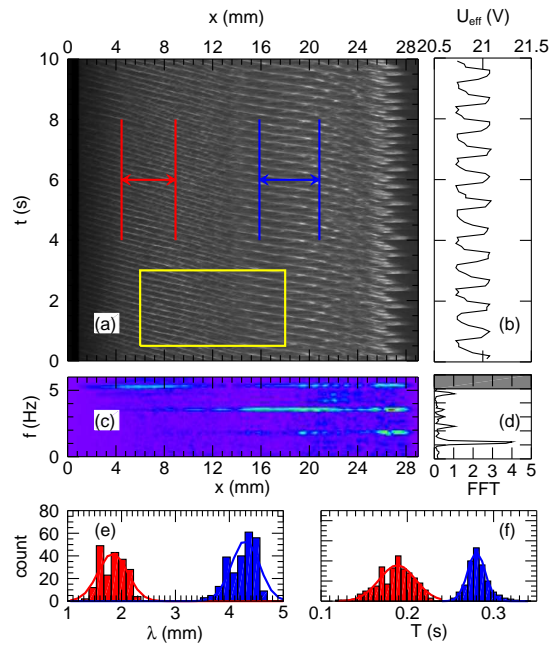}}
\caption{(Color online) (a) Periodgram of horizontally propagating waves from the void to the edge of particle cloud. (b) Evolution of effective voltage. (c) Spectrum distribution along $x$-axis obtained from FFT applied on the periodgram. (d) Spectrum of the effective voltage. Histograms of wavelength $\lambda$ (e) and period $T$ (f) determined from the periodgram: Red histograms (left) correspond to the big particles [red (left) region in (a)] while blue ones (right) correspond to the small particles [blue (right) region in (b)]. The curves in the histograms are Gaussian fits to the distribution. The interface is located at $x\approx13$~mm.
\label{figure2}}
\end{figure}

The waves were excited as the neutral gas pressure was lowered below a critical value ($15$~Pa) \cite{Schwabe:2007}. As we see in Fig.~\ref{figure1}(b), at the pressure of $10$~Pa the particle cloud was compressed vertically by the expansion of the sheath, and the waves propagated through the entire cloud from right to left. The wave crests had a convex shape due to the configuration of the 3D particle cloud. In this letter, we focus on the waves in the region of interest (ROI) with a vertical width of $2$ mm [marked by a blue square in Fig.~\ref{figure1}(b)] around the middle plane where the wave front can be approximated as flat plane. The average number density of the microparticles can be estimated by the pair correlation function. The mean value of the effective current was about $14$~mA and that of the effective voltage ($U_{eff}$) was about $21$~V with an oscillation frequency of $1.1$~Hz, see Fig.~\ref{figure2}(b,d). The plasma parameters were not measured directly but could be estimated based on the published results \cite{Klindworth:2007,Thomas:2008,Heidemann:2011}. The estimated electron density is $n_e\sim1.3\times10^9$~cm$^{-3}$, the electron temperature is $T_e\sim3$~eV and the ion temperature $T_i$ equals to the room temperature. Based on the plasma parameters, we estimate the particle charge and the sound speed for both small and big particles based on three theoretic models \cite{Morfill:2006,Khrapak:2005,Bonitz:2010}. The results are listed in Table.~\ref{table1}.

\begin{figure}[!ht]
\centerline{\includegraphics[bb=100 410 340 500, width=0.45\textwidth]{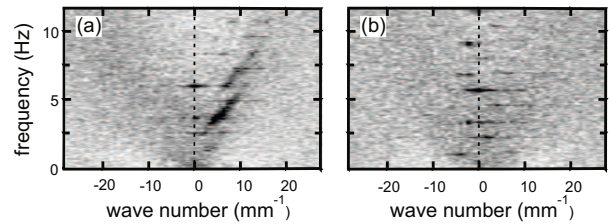}}
\caption{Highly anisotropic particle velocity fluctuation spectra of the binary complex plasma. Panels (a) and (b) show the oscillations in the big and small particle sub-clouds, respectively. Positive wave numbers correspond to the wave propagation to the left in Fig.~\ref{figure1}. Note the presence of sound wave traces (continuum spectrum) and heartbeat oscillation harmonics (discrete spectrum).
\label{figure3}}
\end{figure}

As we can see in Fig.~\ref{figure1}(b), the interface smeared out as the waves developed. However, the position of the interface can still be extrapolated during the process of pressure decrease. Obviously, the wavelength $\lambda$ is not constant over the entire dust cloud. The right part has a larger wavelength than the left part. This is clear in the periodgram in Fig.~\ref{figure2}(a) (see also the video in supplemental materials). The plot takes $500$ consecutive images registered from the experiment. Each vertical column was obtained by averaging over the vertical axis of the ROI in one frame [see the area enclosed by blue square in Fig.~\ref{figure1}(b)] \cite{Schwabe:2007}. Clearly the periodgram is divided into left and right parts, corresponding to the big and small particles. By linearly fitting the wave crests (bright pixel clusters) for the two parts separately, we obtain histograms of the wave length and period from the spatial resp. temporal distances of the wave crests, shown in Fig.~\ref{figure2}(e,f). We determine the mean values of the data by fitting a robust Gaussian to the smoothed histogram. For the waves propagating in the small particles, we have wavelength $\lambda_s = 4.2 \pm 0.3$~mm, wave period $T_s = 0.28 \pm 0.01$~s, frequency $f_s = 3.6 \pm 0.1$~Hz, and phase velocity $\nu_s = 14.2 \pm 1.0$~mm/s. For big particles, we have $\lambda_b = 1.8 \pm 0.3$~mm, $T_b = 0.19 \pm 0.02$~s, $f_b = 5.4 \pm 0.5$~Hz, and $\nu_b = 9.9 \pm 1.8$~mm/s.

The fluctuation spectrum of the binary complex plasma is shown in Fig.~\ref{figure3}. The discrete spectrum suggests the presence of heartbeat oscillation harmonics, while the waves in the continuum are identified as the dust acoustic waves \footnote{Note that Fig.~\ref{figure3} shows the spectrum with both positive and negative wave numbers. In principle the waves can propagate in both directions \cite{Nunomura:2002}.} \cite{Zhdanov:2010}.

The characteristic frequencies of the self-excited waves can also be obtained by directly applying Fast Fourier Transformation (FFT) on the image intensity evolution of the periodgram (representing the number density). The results are shown in Fig.~\ref{figure2}(c). For big particles ($x<10$~mm), we see one characteristic frequency of $5.4$~Hz, in agreement with the analysis of the periodgram. At the interface ($x\approx13$~mm), the peak at $5.4$~Hz is suppressed while another characteristic peak at $3.6$~Hz starts to emerge. The later corresponds to the eigenfrequency of small particles. Inside the sub-cloud of small particles ($x>16$~mm), one sees both peaks at $3.6$~Hz and $5.4$~Hz. For the edge of the void, both the fundamental frequency at $1.8$~Hz and its harmonics are visible.

\begin{figure}[!ht]
\centerline{\includegraphics[bb=0 20 500 580, width=0.4\textwidth]{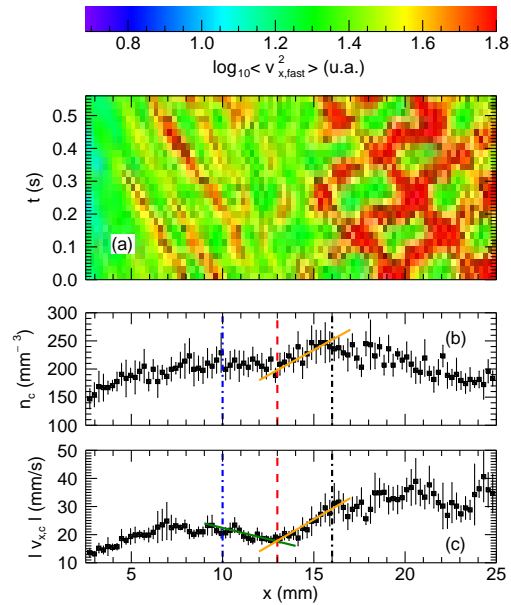}}
\caption{(Color online) Kinetic periodgram (a). The number density $n_c$ (b) and speed $|v_{x,c}|$ (c) at the wave crests along the $x$-axis. The interface is marked by vertical red dashed line and the two characteristic regions close to the interface are marked by vertical blue (left) and black (right) dash-dotted lines. The green [left in (c)] and orange [right in (b,c)] solid lines guide the eyes.
\label{figure4}}
\end{figure}

To further study the dynamics of individual particles in the waves, we plot the kinetic periodgram in Fig.~\ref{figure4}(a), in which the intensity is logarithmly proportional to the mean square speed of the most mobile particle fraction (fastest $25\%$). Because of the limited frame rate of the cameras in the PK-3 Plus laboratory on board the ISS, particle velocity cannot be derived directly from particle tracking for such fast moving and dense particle clouds. However, we can still estimate the particle speed by dividing the length of the elongated shape of individual particles [see the inset in Fig.~\ref{figure1}] by the exposure time of each image. Considering the frequencies of waves of small and big particles, we select the least common multiple of the periods ($\approx 0.6$~s) as the duration to perform an averaging process over $3$~s for better visualization \footnote{The time span covers two periods for small particles and three periods for big particles.}. For big particles, the kinetic periodgram exhibits a left-wards propagating feature with a frequency of $5.4$~Hz, corresponding to the result of FFT analysis. However, for small particles we see a cross-shaped trellis. The left-wards propagating wave has a frequency of $3.6$~Hz, while the right-wards propagating pulses have a frequency of $5.4$~Hz, which is already exhibited in the FFT map in Fig.~\ref{figure2}(c). The maximal particle speed in the right-wards propagating pulses can reach $\sim35\pm5$~mm/s while that in the left-wards propagating waves is $\sim27\pm4$~mm/s.

Several effects contribute to the origin of the waves: There is a heartbeat vibration visible in the electric signals [$1.1$~Hz, see Fig.~\ref{figure2}(b,d)]. The particle movement in the cloud at this frequency is shifted by friction ($\approx 10$~s$^{-1}$), so that only harmonics of $1.8$~Hz are excited as waves in the particle cloud. In the small particle sub-cloud close to the void, the waves excited at approximately $3.6$~Hz also show reversed propagating pulses with a frequency of $5.4$~Hz. The later pulses correspond to the contracting phase of the heartbeat and suggests that this wave was excited by the heartbeat instability. In the big particle sub-cloud, the waves didn't transmit through the interface [suppression of $5.4$~Hz peak at the interface, shown in Fig.~\ref{figure2}(c)]. This suggests that the wave was self-excited by the two-stream instability. However, the heartbeat (generally affecting the entire dust cloud  \cite{Mikikian:2007,Heidemann:2011}) may synchronize the frequency to its third harmonics.

\begin{figure}[!ht]
\centerline{\includegraphics[bb=0 20 500 580, width=0.4\textwidth]{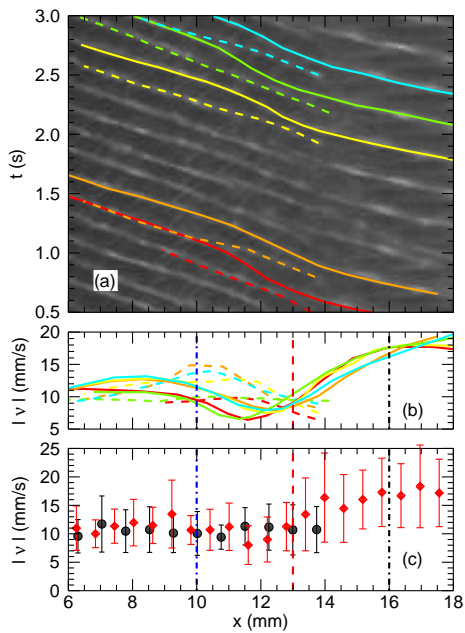}}
\caption{(Color) Periodgram close to the interface (a) and phase speed $|\nu|$ for five typical wave crest pairs (b) and all crests (c). The transmitting wave crests appear in solid curves in (a,b) and in red symbols (c) and self-excited wave crests appear in dashed curves in (a,b) and in black symbols (c). Colors in (a) and (b) correspond to each other. The vertical dashed lines and dash-dotted lines are the same as in Fig.~\ref{figure4}.
\label{figure5}}
\end{figure}

We measured the particle number density $n_c$ and absolute speed $|v_{x,c}|$ at wave crests (ignoring those at wave throughs) and plot them along the $x$-axis in Fig.~\ref{figure4}(b,c). This number density is relatively low close to the void and the edge \footnote{For small particles the number density $n_c$ at wave crests is comparable with the average value. However, for big particles $n_c$ is much greater. This indicates that particle concentration is much lower in the wave throughs in big particles than in small particles.}. As we can see in the figure, $n_c$ is lower in big particles than in small particles. The drop starts at $x\approx16$~mm and reaches a stable level at $x\approx13$~mm, where the interface is roughly located (marked by the red dashed line). We highlight this drop by the orange solid line. For the absolute speed of particles, we see a distinct valley at the interface, where the left shoulder starts at $x\approx10$~mm (blue dash-dotted line) and right shoulder starts at $x\approx16$~mm (black dash-dotted line).

To better understand the subtle change of frequency and wavelength, we zoom into the area close to the interface marked by a yellow square in Fig.~\ref{figure2}(a) and show the details in Fig.~\ref{figure5}(a). At $x\approx16$~mm the wave crests start to bend and we can see the emergence of a new wave front between two existing wave crests at $x\approx13$~mm. We highlight five typical wave crest pairs by different colors in Fig.~\ref{figure5}. The transmitting wave crests from small particles to big particles appear in solid curves while the self-excited wave crests starting at the interface appear in dashed curves. Due to the ``collision'' of the wave crests (with high concentration of heavily charged particles), the phase speed $|\nu|$ decreases for the transmitting waves. Accordingly, $|\nu|$ for the self-excited waves increases, as clearly demonstrated in Fig.~\ref{figure5}(b). As both waves propagate towards the cloud edge, they can either merge to one wave crest (orange-red and cyan-green) or coexist and propagate further (green-yellow). This explains the transition of the frequency of $3.6$~Hz in small particles to $5.6$~Hz in big particles.

To gain better statistics, we tracked all the wave crests close to the interface in Fig.~\ref{figure2}, calculated the phase speed and show the results with errors ($1\sigma$ deviation) in Fig.~\ref{figure5}(c). The drop of the phase speed of the transmitting wave is still clearly visible while the collision of the waves crests smears out due to the randomness of the merge. Combining with the analysis on $n_c$ and $v_{x,c}$, we are able to define the region $13$~mm~$< x < 16$~mm as ``collision zone'' and the region $10$~mm~$< x < 13$~mm as ``merger zone''.

In summary, we have presented the first experimental realization of wave propagation across an interface in complex plasmas. The experiments were performed in a binary complex plasma under microgravity conditions on board the ISS. The small particles and big particles were phase separated with an interface in between due to spinodal decomposition and the difference of electric force and ion drag force. For the big particles, waves were self-excited by the two-stream instability, while for small particles, they were excited by heartbeat instability with the presence of reversed propagating pulses of a different frequency. By studying the dynamics of wave crests at the interface, we recognize a ``collision zone'' and a ``merger zone'' before and after the interface, respectively. The presented results provide a generic picture of wave-wave interaction at the interface between two ``mediums'' and may be exceptionally important for particle-resolved studies of interfacial wave phenomena in future.

\begin{acknowledgments}
The authors acknowledge support from the National Natural Science Foundation of China (NNSFC), Grant No. 11405030 and 11375042. The PK-3 Plus project was funded by the space agency of the Deutsches Zentrum f\"ur Luft- und Raumfahrt eV with funds from the Federal Ministry for Economy and Technology according to a resolution of the Deutscher Bundestag under grant number 50 WP 1203. It was also supported by Roscosmos. A. M. Lipaev, and V. I. Molotkov participated in the PK-3 Plus Project since the predevelopment phase. In particular, they made major contributions to experiment development, planning and conductance. We thank M. Rubin-Zuzic for valuable discussions.
\end{acknowledgments}

\bibliography{waves}




\end{document}